\documentclass{PoS}

\usepackage{amsmath,amssymb}% for dealing with mathematics,
\usepackage{cite}% for dealing with citations
\usepackage{booktabs}
\usepackage{graphicx}
\usepackage{subfigure}
\usepackage{bm}

\newcommand{\mcl}{\mathcal}

\newcommand{\f}[2]{\frac{d^#2 #1}{(2\pi)^#2}}
\newcommand{\vev}[1]{\langle #1 \rangle} %the vacuum expectation value 

\graphicspath{{./figs/}}

\title{Thermodynamics of hadrons using the Gaussian functional method in the linear sigma model}

\ShortTitle{Thermodynamics of hadrons using the Gaussian functional method in the linear sigma model}

\author{\speaker{Shotaro Imai}\\
        Department of Physics, Graduate School of Science, Kyoto University, \\
  Kitashirakawa-oiwake, Sakyo, Kyoto 606-8502, Japan\\
        E-mail: \email{imai@ruby.scphys.kyoto-u.ac.jp}}

\author{Hua-Xing Chen$^a$, Hiroshi Toki$^b$ and Li-Sheng Geng$^a$\\
       \llap{$^a$}School of Physics and Nuclear Energy Engineering and International Research Center\\ 
for Nuclei and Particles in the Cosmos, Beihang University, \\
Beijing 100191, China\\
\llap{$^b$}Reseach Center for Nuclear Physics, Osaka University,\\
Ibaraki, Osaka 567-0047, Japan}%\\
\abstract{
We investigate thermodynamics of hadrons using the Gaussian functional method (GFM) at finite temperature. Since the interaction among mesons is very large, we take into account fluctuations of mesons around their mean field values using the GFM. We obtain the ground state energy by solving the Schr\"{o}dinger equation. The meson masses are obtained using the energy minimization condition.  The resulting mass of the Nambu-Goldstone boson is not zero even in the spontaneous chiral symmetry broken phase due to the non-perturbative effect. We consider then the bound state of mesons using the Bethe-Salpeter equation and show that the Nambu-Goldstone theorem is recovered. We investigate further the behavior of the meson masses and the mean filed value as functions of temperature for the cases of chiral limit and explicit chiral symmetry breaking.}

\FullConference{XV International Conference on Hadron Spectroscopy-Hadron 2013\\
		4-8 November 2013\\
		Nara, Japan }

\begin{document}

\section{Introduction}
The linear sigma model is often used to discuss chiral symmetry breaking and the appearance of Nambu-Goldstone (NG) bosons in the mean field approximation. 
%The interaction among mesons, however, is very large. The fluctuations of mesons around their mean field values should be included in a beyond mean field theory.
The interaction among mesons, however, is very large, and hence the fluctuations of mesons around their mean field values should be included.
 We use the Gaussian functional (GF) method~\cite{Barnes1980} as a non-perturbative many body theory.
 The ground state energy is obtained by solving the Schr\"{o}dinger equation with the trial Gaussian wave functional in the GF method.  %This ground state energy completely corresponds to the CJT effective potential. 
The gap equations for the field vacuum expectation values and the dressed masses are determined by the minimization condition of the energy. This procedure is equivalent to the Hartree-Fock approximation for bosons.
The resulting mass (we call it the dressed mass) of the Nambu-Goldstone boson is, however, not zero, even the chiral symmetry is spontaneously broken in the chiral limit.
 This problem was studied by several authors in the GF framework~\cite{Dmitrasinovic1996,Nakamura2001}.
 In order to recover the Nambu-Goldstone theorem, we have to solve the Bethe-Salpeter (BS) equation. The masses determined by the BS equation are called the physical masses.
 This fact corresponds to the recovery of the translational invariance for nuclear many-body systems using the Random Phase approximation (RPA) after the Hartree-Fock approximation.

In this paper we shall study hadron properties at finite temperature using the BS equation after solving the GF equations.
% to demonstrate the Nambu-Goldstone theorem in the chiral limit.
We show the physical mass spectra for sigma and pion at finite temperature for the two cases of the chiral limit and the explicit chiral symmetry broken. 
The physical sigma meson mass is fixed at 500 MeV in both cases at zero temperature. The physical pion mass stays zero in the chiral symmetry breaking phase in the chiral limit and 138 MeV in the explicit breaking case until the phase transition temperature. When the chiral symmetry is restored, these masses suddenly jump and coincide with the dressed masses.

%In this paper we shall study hadron properties at finite temperature using the BS equation after solving the GF equations to demonstrate the Nambu-Goldstone theorem in the chiral limit. We show the mass spectra for sigma and pion at finite temperature for the two cases of chiral limit and explicit chiral symmetry breaking.

% We show how the chiral symmetry recovers at finite temperature and the behavior of sigma and pion masses for the two cases of chiral limit and explicit chiral symmetry breaking.

\section{The linear sigma model in the Gaussian functional method}
\begin{figure}
\begin{center}
 \subfigure[One-point Schwinger-Dyson equation]{\includegraphics[width=.45\textwidth]{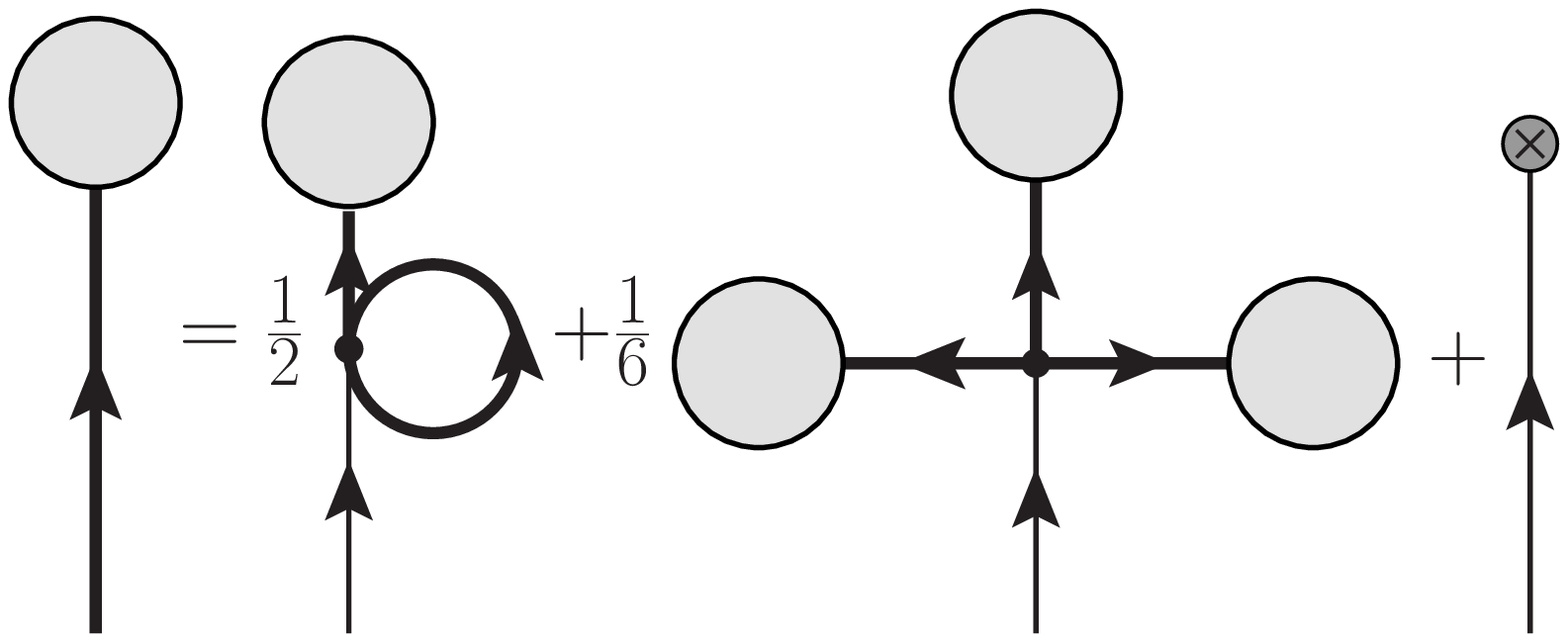}\label{one}}
 \hfill
 \subfigure[Two-point Schwinger-Dyson equation]{\includegraphics[width=.45\textwidth]{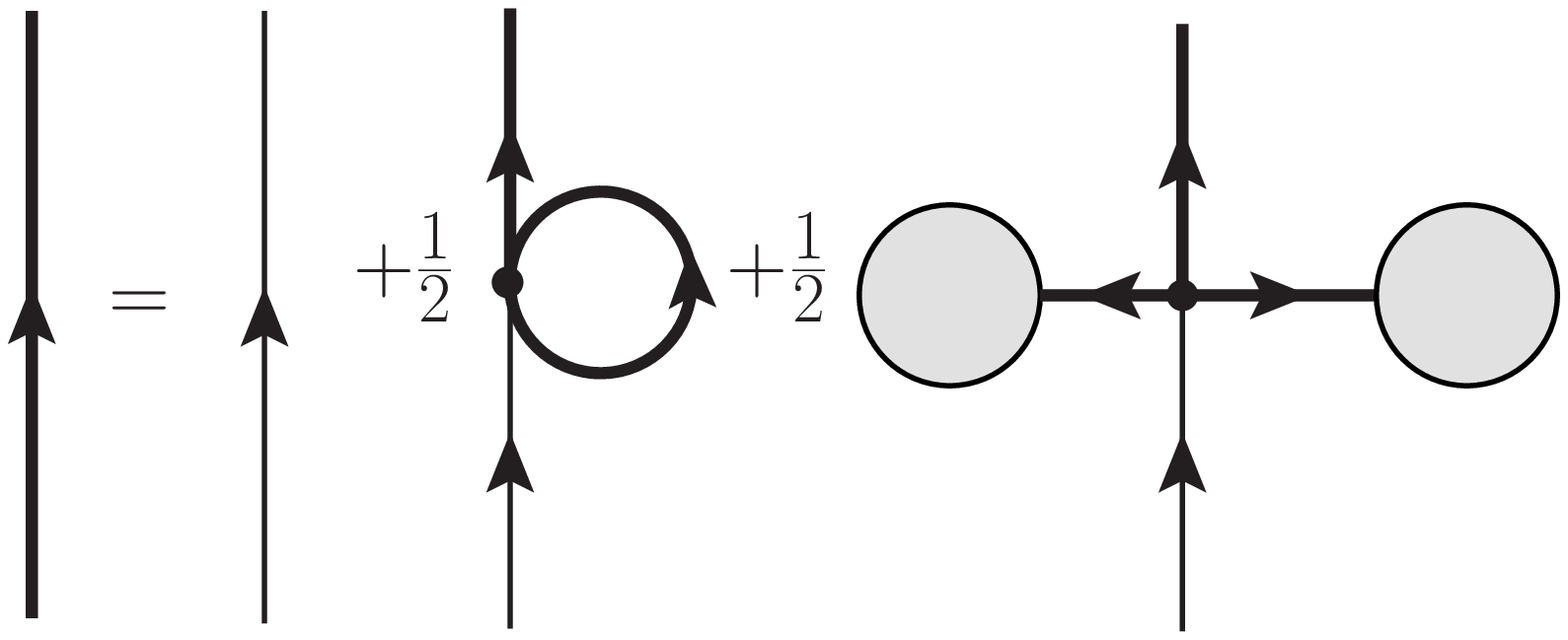}\label{two}}
\caption{The diagramatic representation for one- and two-point Schwinger-Dyson equation. The thin solid line denotes the bare meson fields, the bold solid line is the dressed meson fields, the shaded blob together with the bold line is the vacuum expectation values of the field, and the solid dot in the interaction of the four lines denotes the bare four-point vertex. The diagrams are explicitly multiplied by their symmetry factors.}\label{sd}
\end{center}
\end{figure}

We start from the $O(4)$ symmetric linear sigma model
\begin{align}
 \mcl{L}=&\frac{1}{2}(\partial_\mu\bm{\phi})^2+\frac{1}{2}\mu_0^2\bm{\phi}^2-\frac{\lambda_0}{4}(\bm{\phi}^2)^2+\varepsilon\sigma,\label{sigma}
\end{align}
where the fields are denoted as the column vector $\bm{\phi}=(\phi_0,\phi_1,\phi_2,\phi_3)=(\sigma,\bm{\pi})$.
The parameters in the Lagrangian are the mass $\mu_0$ and the coupling constant $\lambda_0$.
  The explicit chiral symmetry breaking term can be expressed as $\varepsilon\sigma$.
% \begin{align}
%  \mcl{L}_{\chi SB}=-\mcl{H}_{\chi SB}=\varepsilon\sigma.
% \end{align}
We discuss both cases of the chiral limit $\varepsilon\to0$ and the explicit chiral symmetry breaking $\varepsilon=1.86\times10^{6}$ MeV$^3$, which is determined to reproduce the physical poion mass as 138 MeV.
%The mean field approximation does not teke into account radiative corrections of meson loops. 
The Gaussian functional method (GFM) can take into account radiative corrections of meson loops.
We use the following Gaussian ground state wave functional ansatz:
\begin{align}
\Psi[\bm{\phi}]=\mcl{N}\exp\left(-\frac{1}{4\hbar}\int d\bm{x} d\bm{y} [\phi_i(\bm{x})-\vev{\phi_i(\bm{x})}]G^{-1}_{ij}(\bm{x},\bm{y})[\phi_j(\bm{y})-\vev{\phi_j(\bm{y})}]\right),
\end{align}
where $\vev{\phi_i}$ is the vacuum expectation value of the $i$-th field, and we define
\begin{align}
G_{ij}(\bm{x},\bm{y})=\frac{1}{2}\delta_{ij}\int \f{{\bf k}}{3}\frac{1}{\sqrt{\bm{k}^2+M_i^2}}e^{i\bm{k}\cdot(\bm{x}-\bm{y})},
\end{align}
where $M_i$ is the ``dressed'' mass of the $\sigma$ meson (denoted by $M_0=M_{\sigma}$) and the $\pi$ meson ($M_{1,2,3}=M_{\pi}$). 
%The ``dressed'' means we still have to solove the Bethe-Salpeter equation to calculate the physical masses. 
The ground state energy is defined as
\begin{align}
 \mcl{E}(M_{i},\vev{\phi_i})=\int\mcl{D}\bm{\phi}\Psi^*[\bm{\phi}]\mcl{H}[\bm{\phi}]\Psi[\bm{\phi}],
\end{align}
where the Hamiltonian $\mcl{H}$ is defined by the Lagrangian~\eqref{sigma} through the Legendre transformation.
The variational parameters $\vev{\phi_i}$ and $M_i$ are determined by the energy minimization condition,
\begin{align}
 \left(
\frac{\partial {\cal E}(M_{i}, \langle  \phi_{i} \rangle)}{\partial \langle \phi_{i} \rangle,M_i}
\right)_{\text{min}} = 0 \, ,
\mbox{ for } i = 0 \ldots 3, \label{min}
\end{align}
which is equivalent to the one- and two-point Schwinger-Dyson euqations, shown in Fig~\ref{sd}.
The condition~\eqref{min} leads to the mean field values and the dressed masses as 
%which leads to
\begin{align}
 \vev{\phi_0}=&v,\quad \vev{\phi_i}=0 \, \text{ for } \, i=1,2,3\\
  \mu_{0}^{2}=&-\frac{\varepsilon}{v}+{\lambda_{0}}\left[v^{2}+3I_{0}(M_\sigma)+3I_{0}(M_\pi)\right],\\
 M_\sigma^{2}%=&-\mu_{0}^{2}+{\lambda_{0}}\left[3v^{2}+3I_{0}(M_\sigma)+3I_{0}(M_\pi)\right]\nn
=&\frac{\varepsilon}{v}+2\lambda_0 v^2,\\
  M_\pi^{2}%=&-\mu_{0}^{2}+{\lambda_{0}}\left[v^{2}+I_{0}(M_\sigma)+5I_{0}(M_\pi)\right]\nn
=&\frac{\varepsilon}{v}+2\lambda_0\left[I_0(M_{\pi})-I_0(M_{\sigma})\right],\label{dmp}
\end{align}
where the loop integration, 
\begin{align}
 I_0(M_i)=\frac{1}{2}\int_{0}^{\Lambda}\f{\bm{k}}{3}\frac{1}{\sqrt{\bm{k}^2+M_{i}^2}}
%=i\int^{\Lambda}\f{k}{4}\frac{1}{k^2-M_i^2+i\epsilon},
\end{align}
and the momentum cutoff $\Lambda$ have been introduced. Although the pion mass $M_{\pi}$ should be zero, when the chiral symmetry is broken in the chiral limit, it is finite due to the non-perturbative loop correction (see Eq.~\eqref{dmp}). We cannot identify the ``dressed'' pion as the NG boson.% and the Bethe-Salpeter equation provies the physical particles.

\section{The Bethe-Salpeter equation}
\begin{figure}[htbp]
\begin{center}
\subfigure[The Bethe-Salpeter equation with the T-matrix and the interaction kernel.]{\includegraphics[width=0.7\columnwidth]{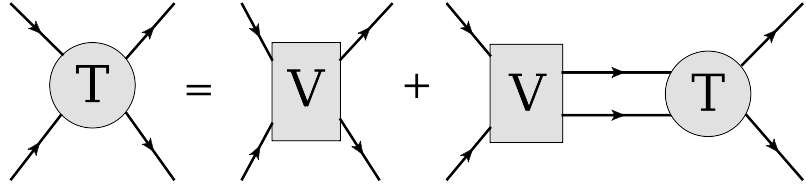}}
\subfigure[The interaction kernel entering the Bethe-Salpeter equation. The shaded blob together with the bold line is the vacuum expectation value of the field.]{\includegraphics[width=0.5\columnwidth]{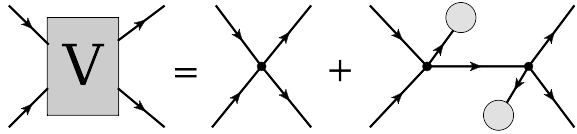}}
\caption{The diagramatic expression of the Bethe-Salpeter equation. All lines represent the dressed fields. The physical mass is obtained from the pole of the T-matrix.}\label{bs}
\end{center}
\end{figure}

\begin{figure}
\begin{center}
 \subfigure[The chiral limit case ($\varepsilon=0$)]{\includegraphics[width=.45\textwidth]{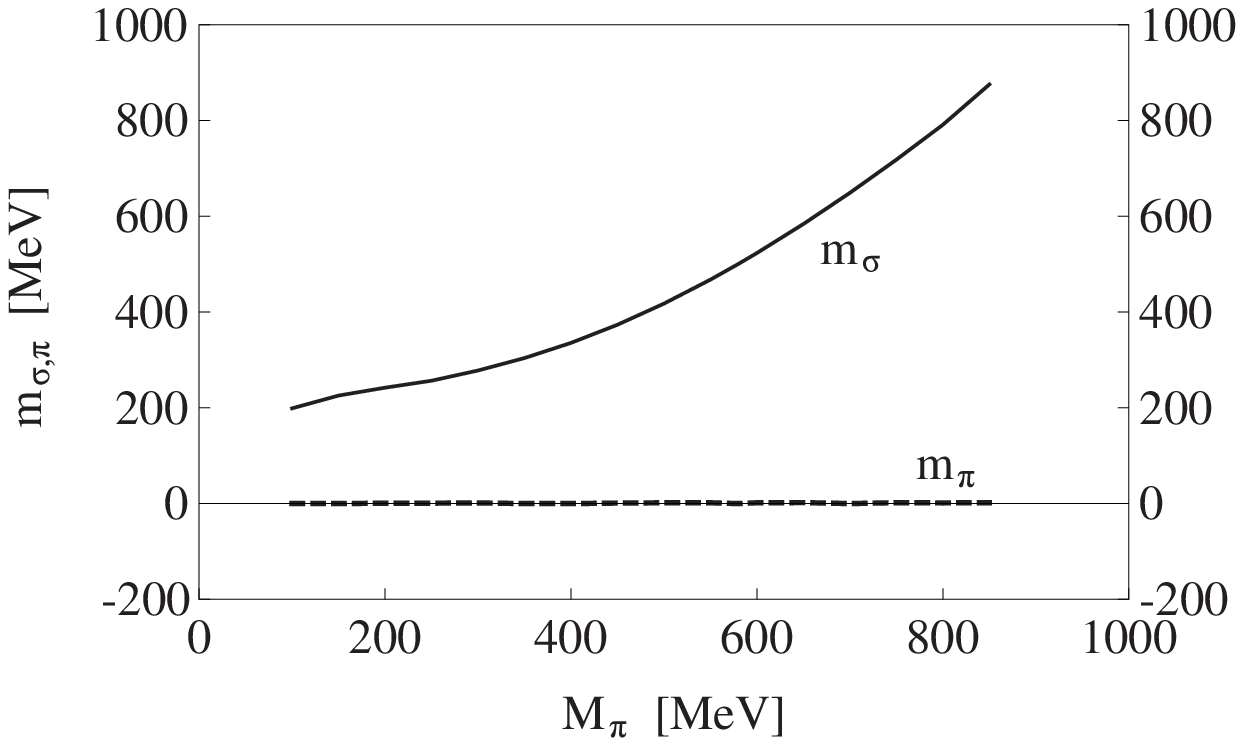}\label{bsc}}
 \subfigure[The explicit chiral symmetry breaking case ($\varepsilon\neq0$)]{\includegraphics[width=.45\textwidth]{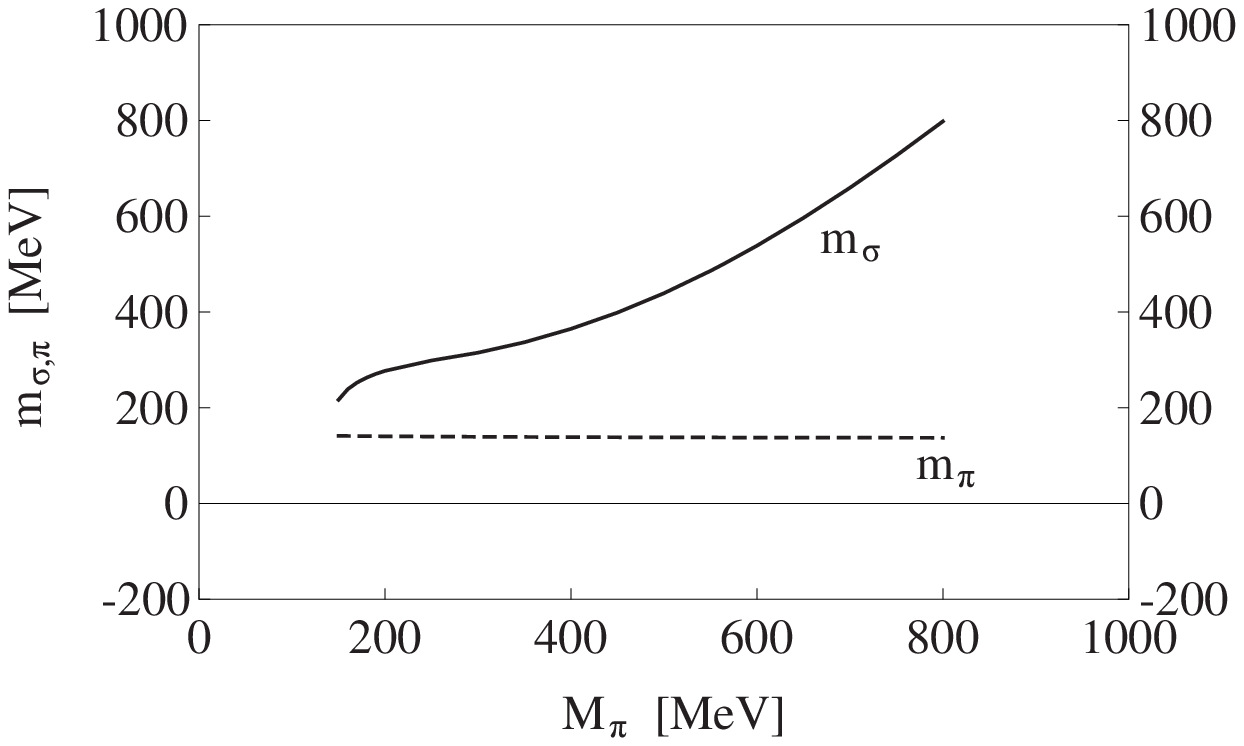}\label{bsb}}
 \caption{The physical masses $m_{\sigma}$ and $m_{\pi}$ as a function of the dressed pion mass $M_{\pi}$.}
\end{center}
\end{figure}

The Physical mass can be obtained by solving the Bethe-Salpeter equation.
 The diagramatic expression is shown in Fig. \ref{bs}.
 The single channel Bethe-Salpeter equation of the $\sigma$-$\pi$ scattering gives the physical pion mass $m_\pi$ and the coupled-channels of $\sigma$-$\sigma$ and $\pi$-$\pi$ scatterings give the physical $\sigma$ mass $m_\sigma$.
 %We note that by doing this the Nambu-Goldstone theorem can be fulfilled in the chiral limit under proper regularization Eq.~(\ref{eq:regularization})~\cite{nakamura01,chen10}.
We show the result of the physical masses $m_{\sigma,\pi}$ as a function of the dressed pion mass $M_{\pi}$ in the chiral limit in Fig.~\ref{bsc}, and the explicit breaking case in Fig.~\ref{bsb}. The Nambu-Goldstone theorem is always fulfilled in the chiral limit for any value of the dressed pion mass $M_{\pi}$. 
%Hence, the physical masses should be considered as a bound state of the dressed particles.
 In the explicit symmetry breaking case $\varepsilon\neq0$, the physical pion mass is fixed at $m_{\pi}=138$ MeV.

\section{Thermodynamics of hadrons in the GFM+BS scheme}
\begin{figure}
\begin{center}
 \subfigure[The chiral limit case ($\varepsilon=0$)]{\includegraphics[width=.45\textwidth]{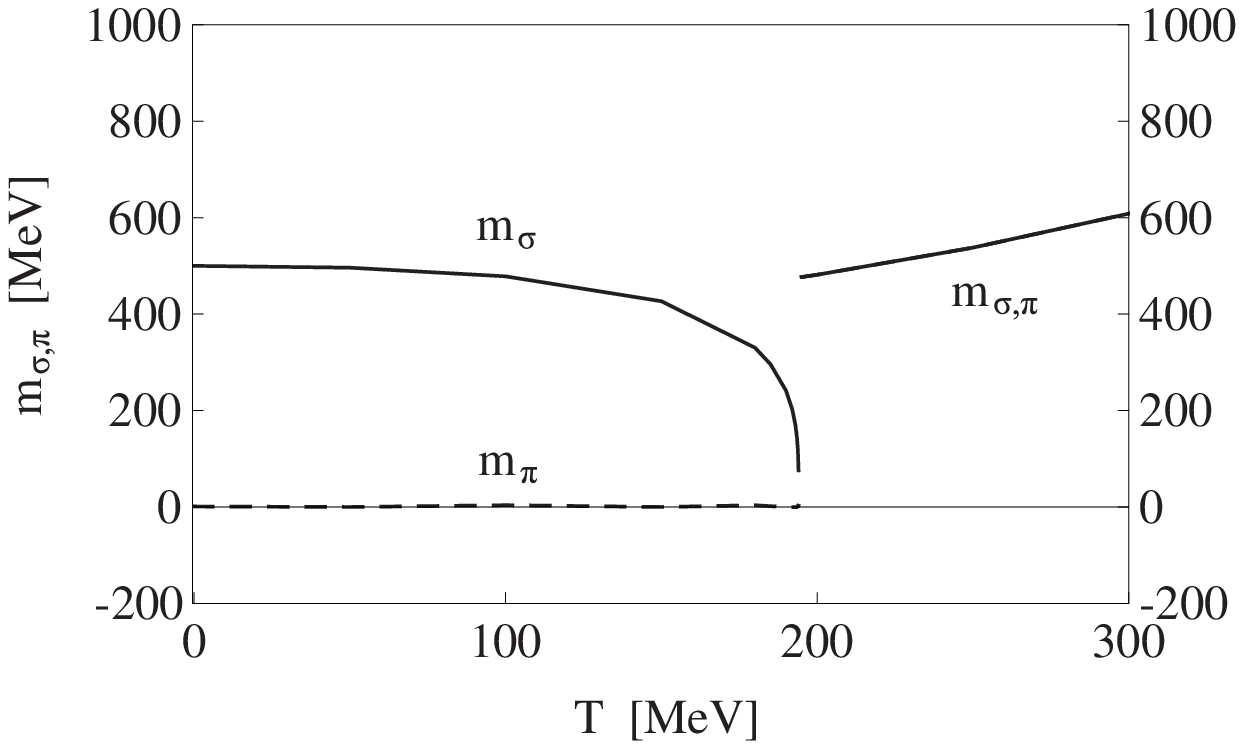}\label{tbsc}}
 \subfigure[The explicit chiral symmetry breaking case ($\varepsilon\neq0$)]{\includegraphics[width=.45\textwidth]{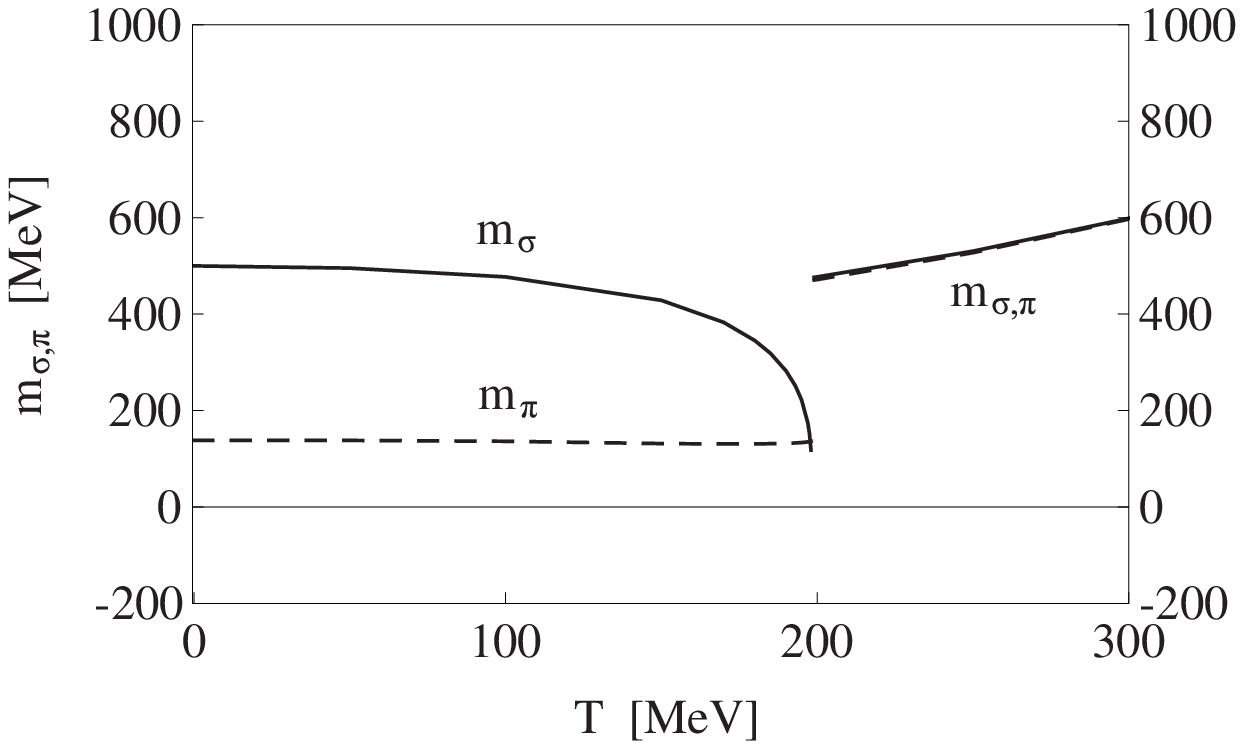}\label{tbsb}}
 \caption{The physical masses $m_{\sigma}$ and $m_{\pi}$ as a function of temperature.}\label{tbs}%The masses coincide above the critical temperature $T_c$.
\end{center}
\end{figure}

In this section we discuss the mass spectra at finite temperature.
 The parameters of this theory are fixed to reproduce the pion mass which is zero in the chiral limit and 138 MeV in the symmetry breaking case, the pion decay constant $f_{\pi}=93$ MeV and the sigma meson mass $m_{\sigma}=500$ MeV at zero temperature.
The result is shown in Fig.~\ref{tbs}. The physical pion mass $m_{\pi}$ stays at zero or 138 MeV until the critical temperature $T_c$, which means that the chiral symmetry is spontaneously broken. The masses suddenly jump and coincide with each other above $T_c$, since the chiral symmetry is restored.
By solving the Bethe-Salpeter equation, we can see that the physical masses coincide with the dressed masses when the chiral symmetry is restored ($v\to0$).
This behavior suggests that the physical states are bound states of the dressed mesons in the symmetry broken phase, and the physical states become the dressed mesons in the symmetric phase.
%\textcolor{red}{the bound states decay into the dressed mesons in the chiral symmetric phase}.
% The list of the parameters are shown in Table \ref{param}.
% \begin{table}
% \caption{The parameter list of the theory.} \label{param}
% \begin{center}
%   \begin{tabular}{c|cccc}
% \toprule
%  &$\varepsilon$ [MeV$^3$] & $\mu_0$ [MeV] & $\lambda_0$ & $\Lambda$ [MeV]\\
% \midrule
%  chiral limit & 0 & 1680 &83.6&800\\
%  explicit breaking &1.86$\times$10$^6$&1610&75.5&800\\
% \bottomrule
%  \end{tabular} 
% \end{center}
% \end{table}

\section{Summary}
We have investigated the meson mass spectra at finite temperature using the Gaussian functional method with the Bethe-Salpeter equation in the linear sigma model. 
%The fluctuations around the mean field values are included in the GFM.
% However, the Nambu-Goldstone theorem cannot be fullfilled at this stage even the chiral symmetry is spontaneously broken. Then we have to consider the bound state of the partcles by the Bethe-Salpeter equation. 
The Nambu-Goldstone theorem is fulfilled in this scheme, which is explicitly shown in this study. We discuss the thermodynamical properties of the mesons.
The physical states can be considered as the bound states of the dressed mesons in the chiral symmetry broken phase.
% The bound states decay into the dressed fields in the chiral symmetric phase.

\acknowledgments{This work is supported by the JSPS research grant (S): 21540267.  H.X.C is supported by the National Natural Science Foundation of China under Grant No.\,11205011, and the Fundamental Research Funds for the Central Universities.  S.I is supported by the Grant for Scientific Research [Priority Areas ``New Hadrons'' (E01:21105006), (C) No.\,23540306] from the Ministry of Education, Culture,Science and Technology (MEXT) of Japan. L.S.G is supported by the National Natural Science Foundation of China under Grant No.\,11005007.}

%  \bibliographystyle{ws-procs9x6}
% \bibliography{./bib/njl,./bib/mynotes,./bib/sigma,/bib/thermodynamics,./bib/text,}

\end{document}